\documentclass[aps,pra,superscriptaddress,reprint]{revtex4-1}
\usepackage{amsmath,amsthm}
\usepackage{amsfonts}
\usepackage[colorlinks=true]{hyperref}
\newtheorem{thm}{Theorem}

\newtheorem{lem}[thm]{Lemma}

\theoremstyle{definition}

\theoremstyle{remark}

\begin{document}
\title{Gaussian states minimize the output entropy of one-mode quantum Gaussian channels}
\author{Giacomo De Palma}
\affiliation{QMATH, Department of Mathematical Sciences, University of Copenhagen, Universitetsparken 5, 2100 Copenhagen, Denmark}
\affiliation{NEST, Scuola Normale Superiore and Istituto Nanoscienze-CNR, I-56126 Pisa, Italy}
\affiliation{INFN, Pisa, Italy}
\author{Dario Trevisan}
\affiliation{Universit\`a degli Studi di Pisa, I-56126 Pisa, Italy}
\author{Vittorio Giovannetti}
\affiliation{NEST, Scuola Normale Superiore and Istituto Nanoscienze-CNR, I-56126 Pisa, Italy}

\begin{abstract}
We prove the longstanding conjecture stating that Gaussian thermal input states minimize the output von Neumann entropy of one-mode phase-covariant quantum Gaussian channels among all the input states with a given entropy.
Phase-covariant quantum Gaussian channels model the attenuation and the noise that affect any electromagnetic signal in the quantum regime.
Our result is crucial to prove the converse theorems for both the triple trade-off region and the capacity region for broadcast communication of the Gaussian quantum-limited amplifier.
Our result extends to the quantum regime the Entropy Power Inequality that plays a key role in classical information theory.
Our proof exploits a completely new technique based on the recent determination of the $p\to q$ norms of the quantum-limited amplifier [De Palma et al., arXiv:1610.09967].
This technique can be applied to any quantum channel.
\end{abstract}

\maketitle

\section{Introduction}
Signal attenuation and noise unavoidably affect electromagnetic communications through metal wires, optical fibers or free space.
Since the energy carried by an electromagnetic pulse is quantized, quantum effects must be taken into account \cite{gordon1962quantum}.
They become relevant for low-intensity signals, such as for satellite communications, where the receiver can be reached by only few photons for each bit of information \cite{chen2012optical}.
In the quantum regime, signal attenuation and noise are modeled by phase-covariant quantum Gaussian channels \cite{chan2006free,braunstein2005quantum,holevo2013quantum,weedbrook2012gaussian,holevo2015gaussian} (sometimes also called gauge-covariant quantum Gaussian channels).

The maximum achievable communication rate of a channel depends on the minimum noise achievable at its output, that is quantified by the output von Neumann entropy \cite{wilde2013quantum,holevo2013quantum}.
We prove in the case of one mode the longstanding constrained minimum output entropy (CMOE) conjecture \cite{guha2007classicalproc,guha2007classical,guha2008entropy,guha2008capacity,wilde2012information,wilde2012quantum} stating that Gaussian thermal input states minimize the output entropy of phase-covariant quantum Gaussian channels among all the input states with a given entropy.

The classical counterpart of the CMOE conjecture states that Gaussian input probability distributions minimize the output Shannon differential entropy of classical Gaussian channels among all the input probability distributions with a given entropy, and it is implied by the Entropy Power Inequality (EPI) \cite{cover2006elements,blachman1965convolution}.
The EPI is fundamental in classical information theory.
It is necessary to prove the optimality of Gaussian encodings for the transmission of information through the classical broadcast and wiretap channels \cite{bergmans1974simple,leung1978gaussian}, and it provides bounds for the information capacities of non-Gaussian classical communication channels \cite{shannon2001mathematical} and for the convergence rate in the Central Limit Theorem \cite{barron1986entropy}.
A quantum generalization of the proof of the EPI permits to prove the quantum EPI (qEPI) \cite{konig2013classical,konig2013limits,konig2014entropy,de2014generalization,de2015multimode}, that provides a lower bound to the output von Neumann entropy of quantum Gaussian channels in terms of the input entropy.
However, the qEPI is \emph{not} saturated by quantum Gaussian states, hence it is not sufficient to prove the CMOE conjecture.
The MOE conjecture has first been proven in a completely different way in the version stating that pure Gaussian input states minimize the output entropy of any phase covariant and contravariant quantum Gaussian channel among all the possible pure and mixed input states \cite{garcia2012majorization,mari2014quantum,giovannetti2015solution,giovannetti2015majorization,holevo2015gaussian}.
This fundamental result has permitted to determine the classical communication capacity of these channels \cite{giovannetti2014ultimate} and to prove that this capacity is additive under tensor product, i.e. it is not increased by entangling the inputs \cite{holevo2015gaussian}.
The CMOE conjecture has then been proven for the one-mode quantum-limited attenuator \cite{de2016gaussian,de2015passive} using Lagrange multipliers.
Unfortunately the same proof does not work in the presence of amplification or noise.

In this Letter we prove the CMOE conjecture for any one-mode phase-covariant quantum Gaussian channel.
This result implies the CMOE conjecture also for one-mode phase-contravariant quantum Gaussian channels (\cite{qi2016thermal}, Section VI).
Our result both extends the EPI to the quantum regime and generalizes the unconstrained minimum output entropy conjecture of \cite{mari2014quantum,giovannetti2015solution,giovannetti2015majorization,holevo2015gaussian}, that has permitted to determine the classical capacity of any phase-covariant quantum Gaussian channel \cite{giovannetti2014ultimate} (see also \cite{schafer2013equivalence}).
Our result is necessary to prove the converse theorems that guarantee the optimality of Gaussian encodings for two communication tasks involving the quantum-limited amplifier \cite{qi2016capacities}.
The first is the triple trade-off coding \cite{wilde2012public}, that allows to simultaneously transmit both classical and quantum information and to generate shared entanglement, or to simultaneously transmit both public and private classical information and to generate a shared secret key.
The second is broadcast communication \cite{yard2011quantum,savov2015classical}, i.e.\ classical communication with two receivers.

Our proof exploits a completely new technique that links the CMOE conjecture to the $p\to q$ norms \cite{holevo2006multiplicativity,holevo2015gaussian}, and is based on the result stating that Gaussian thermal input states saturate the $p\to q$ norms of the one-mode quantum-limited amplifier \cite{de2016pq}.
This technique can be used to determine the minimum output entropy for fixed input entropy for any quantum channel whose $p\to q$ norms are known.

We start presenting quantum Gaussian channels, and then prove the CMOE conjecture.
We refer the reader to Appendix \ref{app} for some technical details.

\section{Bosonic Gaussian systems}
We consider the Hilbert space of one harmonic oscillator, or one mode of the electromagnetic radiation.
The ladder operator $\hat{a}$ satisfies the bosonic canonical commutation relation $\left[\hat{a},\;\hat{a}^\dag\right]=\hat{\mathbb{I}}$, and the Hamiltonian $\hat{N}=\hat{a}^\dag\hat{a}$ counts the number of excitations, or photons.
The density matrix of the thermal Gaussian state with average energy $E\ge0$ is
\begin{equation}\label{eq:omegaz}
\hat{\omega}_E=\sum_{n=0}^\infty \frac{1}{E+1}\left(\frac{E}{E+1}\right)^n|n\rangle\langle n|\;,
\end{equation}
where the Fock states $|n\rangle$  are the eigenvectors of $\hat{N}$.
Its von Neumann entropy is
\begin{equation}\label{eq:Sz}
S\left(\hat{\omega}_E\right)= \left(E+1\right)\ln\left(E+1\right)-E\ln E:=g(E)\;.
\end{equation}

Phase covariant and contravariant quantum Gaussian channels \cite{holevo2007one} are the quantum channels that preserve the set of thermal Gaussian states.
Phase-covariant quantum Gaussian channels are constituted by quantum attenuators, quantum amplifiers,  and additive-noise channels.

The quantum attenuator $\mathcal{E}_{\lambda,E}$ of transmissivity $0\le\lambda\le 1$ and thermal energy $E\ge 0$ mixes the input state $\hat{\rho}$ with the thermal Gaussian state $\hat{\omega}_E$ of an environmental quantum system $B$ through a beamsplitter of transmissivity $\lambda$
(case (C) of \cite{holevo2007one} with $k=\sqrt{\lambda}$ and $N_0=E$):
\begin{equation}
\mathcal{E}_{\lambda,E}\left(\hat{\rho}\right)=\mathrm{Tr}_B\left[\hat{U}_\lambda\left(\hat{\rho}\otimes\hat{\omega}_E\right)\hat{U}_\lambda^\dag\right]\;.
\end{equation}
Here $\mbox{Tr}_B[ \cdots]$ is the partial trace over the environment $B$,
\begin{equation}
\hat{U}_\lambda=\exp\left(\left(\hat{a}^\dag\hat{b}-\hat{a}\,\hat{b}^\dag\right)\arccos\sqrt{\lambda}\right)
\end{equation}
is the unitary operator implementing the beamsplitter, and $\hat{b}$ is the ladder operator of $B$ (see Section 1.4.2 of \cite{ferraro2005gaussian}).
For $E=0$ the state of the environment is the vacuum and the attenuator is quantum-limited.
We put $\mathcal{E}_{\lambda,0}=\mathcal{E}_\lambda$ for simplicity.
The action of the quantum attenuator on thermal Gaussian states is~\cite{holevo2013quantum}
\begin{equation} \label{TRANSFGAUS}
\mathcal{E}_{\lambda,E}\left(\hat{\omega}_{E'}\right)=\hat{\omega}_{\lambda E'+(1-\lambda)E}\;.
\end{equation}

The quantum amplifier $\mathcal{A}_{\kappa,E}$ of amplification parameter $\kappa\ge1$ and thermal energy $E\ge 0$ performs a two-mode squeezing on the input state $\hat{\rho}$ and the thermal Gaussian state $\hat{\omega}_E$ of  $B$ (case (C) of \cite{holevo2007one} with $k=\sqrt{\kappa}$ and $N_0=E$):
\begin{equation}
\mathcal{A}_{\kappa,E}\left(\hat{\rho}\right)=\mathrm{Tr}_B\left[\hat{U}_\kappa\left(\hat{\rho}\otimes \hat{\omega}_E\right)\hat{U}_\kappa^\dag\right]\;,
\end{equation}
where
\begin{equation}\label{eq:defUk}
\hat{U}_\kappa=\exp\left(\left(\hat{a}^\dag\hat{b}^\dag-\hat{a}\,\hat{b}\right)\mathrm{arccosh}\sqrt{\kappa}\right)
\end{equation}
is the squeezing  unitary operator.
Again for $E=0$ the amplifier is quantum-limited and we put $\mathcal{A}_{\kappa,0}=\mathcal{A}_\kappa$ for simplicity.
The action of the amplifier on thermal Gaussian states is
\begin{equation}\label{eq:AE'}
\mathcal{A}_{\kappa,E}\left(\hat{\omega}_{E'}\right)=\hat{\omega}_{\kappa E'+(\kappa-1)(E+1)}\;.
\end{equation}

The additive-noise channel $\mathcal{N}_E$ adds $E\ge0$ to the energy of the input state, and can be expressed as a quantum-limited amplifier composed with a quantum limited attenuator (case ($\text{B}_2$) of \cite{holevo2007one} with $N_c=E$):
\begin{equation}
\mathcal{N}_E=\mathcal{A}_{E+1}\circ\mathcal{E}_\frac{1}{E+1}\;.
\end{equation}
Its action on thermal Gaussian states is
\begin{equation}
\mathcal{N}_E\left(\hat{\omega}_{E'}\right) = \hat{\omega}_{E'+E}\;.
\end{equation}

Any phase-covariant quantum Gaussian channel can be expressed as a quantum-limited amplifier composed with a quantum-limited attenuator \cite{mari2014quantum,giovannetti2015solution,holevo2015gaussian,garcia2012majorization}:
\begin{equation}\label{eq:dec}
\mathcal{E}_{\lambda,E} = \mathcal{A}_{\kappa'}\circ\mathcal{E}_{\lambda'}\;,\qquad \mathcal{A}_{\kappa,E} = \mathcal{A}_{\kappa''}\circ\mathcal{E}_{\lambda''}\;,
\end{equation}
where
\begin{align}
\lambda' &= \frac{\lambda}{\left(1-\lambda\right)E+1}\;,\qquad &\kappa' = \left(1-\lambda\right)E+1\;,\nonumber\\
\lambda'' &= \frac{1}{\left(1-\frac{1}{\kappa}\right)E+1}\;,\qquad &\kappa'' = \kappa\left(\left(1-\frac{1}{\kappa}\right)E+1\right)\;.
\end{align}

The phase-contravariant channel $\tilde{\mathcal{A}}_{\kappa,E}$ is the weak complementary of the amplifier $\mathcal{A}_{\kappa,E}$ (case (D) of \cite{holevo2007one} with $k=\sqrt{\kappa-1}$ and $N_0=E$):
\begin{equation}
\tilde{\mathcal{A}}_{\kappa,E}\left(\hat{\rho}\right)=\mathrm{Tr}_A\left[\hat{U}_\kappa\left(\hat{\rho}\otimes \hat{\omega}_E\right)\hat{U}_\kappa^\dag\right]\;,
\end{equation}
where $\hat{U}_\kappa$ is the two-mode squeezing unitary defined in \eqref{eq:defUk} and where now the partial trace is performed over the system $A$.
The action of $\tilde{\mathcal{A}}_{\kappa,E}$ on thermal Gaussian states is \cite{holevo2013quantum}
\begin{equation}
\tilde{\mathcal{A}}_{\kappa,E}\left(\hat{\omega}_{E'}\right)=\hat{\omega}_{(\kappa-1)(E'+1)+\kappa E}\;.
\end{equation}
\section{Gaussian Optimization}
The CMOE conjecture for the quantum-limited attenuator was proven in Ref.~\cite{de2016gaussian}.
\begin{thm}[CMOE conjecture for the quantum-limited attenuator \cite{de2016gaussian}]\label{thm:SP}
Gaussian thermal input states minimize the output entropy of the one-mode quantum-limited attenuator among all the input states with a given entropy, i.e.\ for any input state $\hat{\rho}$ and any $0\leq\lambda\leq1$
\begin{equation}\label{epni}
S\left(\mathcal{E}_\lambda\left(\hat{\rho}\right)\right)\geq g\left(\lambda\;g^{-1}\left(S\left(\hat{\rho}\right)\right)\right) = S\left(\mathcal{E}_\lambda\left(\hat{\omega}\right)\right)\;,
\end{equation}
where $\hat{\omega}$ is the thermal Gaussian state with $S(\hat{\omega})=S(\hat{\rho})$.
\end{thm}
Here we prove the CMOE conjecture for any phase covariant and contravariant one-mode quantum Gaussian channel.
The first step is the proof for the quantum-limited amplifier.
\begin{thm}[CMOE conjecture for the quantum-limited amplifier]\label{thm:SA}
Gaussian thermal input states minimize the output entropy of the one-mode quantum-limited amplifier among all the input states with a given entropy, i.e.\ for any input state $\hat{\rho}$ and any $\kappa\ge1$
\begin{equation} \label{CLAIM}
S\left(\mathcal{A}_\kappa\left(\hat{\rho}\right)\right)\ge g\left(\kappa\;g^{-1}\left(S\left(\hat{\rho}\right)\right)+\kappa-1\right) = S\left(\mathcal{A}_\kappa\left(\hat{\omega}\right)\right)\;,
\end{equation}
where $\hat{\omega}$ is the thermal Gaussian state with $S(\hat{\omega})=S(\hat{\rho})$.
\begin{proof}
Since $\mathcal{A}_1$ is the identity channel, the claim is trivial for $\kappa=1$.
We can then assume $\kappa>1$.
For $S\left(\hat{\rho}\right)=0$ the claim is implied by the Gaussian minimum output entropy conjecture \cite{giovannetti2015solution}, stating that the vacuum input state minimizes the output entropy of any phase-covariant quantum Gaussian channel among all the possible input states, and in particular among the pure input states.
We can then assume $S\left(\hat{\rho}\right)>0$.
The starting point of our proof is the result of Ref.~\cite{de2016pq} stating that thermal Gaussian states saturate the $p\to q$ norms of the quantum-limited amplifier.
\begin{thm}[$p\to q$ norms of the quantum-limited amplifier \cite{de2016pq}]\label{thm:pq}
For any $1<p<q$ and any $\kappa>1$ the $p\to q$ norm of $\mathcal{A}_\kappa$ is saturated by a thermal Gaussian state $\hat{\omega}$ (that depends on $\kappa$, $p$ and $q$), i.e. for any quantum state $\hat{\rho}$
\begin{equation}\label{eq:pq}
\frac{\left\|\mathcal{A}_\kappa\left(\hat{\rho}\right)\right\|_q}{\left\|\hat{\rho}\right\|_p}\le \frac{\left\|\mathcal{A}_\kappa\left(\hat{\omega}\right)\right\|_q}{\left\|\hat{\omega}\right\|_p}\;,
\end{equation}
where
\begin{equation}
\left\|\hat{X}\right\|_\alpha=\left(\mathrm{Tr}\,\hat{X}^\alpha\right)^\frac{1}{\alpha}\;,\qquad\alpha>1\;,\qquad\hat{X}\ge0
\end{equation}
is the Schatten $\alpha$ norm \cite{schatten1960norm,holevo2006multiplicativity}.
\end{thm}
Let $\hat{\rho}$ be a quantum state with $0<S\left(\hat{\rho}\right)<\infty$, and let $\hat{\omega}$ be the thermal Gaussian state with $S\left(\hat{\rho}\right)=S\left(\hat{\omega}\right)$.

For any $\alpha>1$, the $\alpha$ R\'enyi entropy of a quantum state $\hat{\sigma}$ is
\begin{equation}\label{eq:defSa}
S_\alpha\left(\hat{\sigma}\right)=\frac{\alpha}{1-\alpha}\ln\left\|\hat{\sigma}\right\|_\alpha\;,
\end{equation}
and satisfies \cite{holevo2015gaussian}
\begin{equation}\label{eq:propSa}
S_\alpha\left(\hat{\sigma}\right) \le S\left(\hat{\sigma}\right)\;,\qquad \lim_{\alpha\to 1}S_\alpha\left(\hat{\sigma}\right)=S\left(\hat{\sigma}\right)\;.
\end{equation}

From Lemma \ref{lem} of Appendix \ref{app}, for any $1<q<3/2$ there exists $1<p(q)<q$ such that the $p(q)\to q$ norm of $\mathcal{A}_\kappa$ is saturated by $\hat{\omega}$.
We then have
\begin{align}\label{eq:SAq}
&S\left(\mathcal{A}_\kappa\left(\hat{\rho}\right)\right) \ge S_q\left(\mathcal{A}_\kappa\left(\hat{\rho}\right)\right) = \frac{q}{1-q}\ln\left\|\mathcal{A}_\kappa\left(\hat{\rho}\right)\right\|_q\nonumber\\
&\ge \frac{q}{1-q}\ln\frac{\left\|\mathcal{A}_\kappa\left(\hat{\omega}\right)\right\|_q\left\|\hat{\rho}\right\|_{p(q)}}{\left\|\hat{\omega}\right\|_{p(q)}}\nonumber\\
&= S_q\left(\mathcal{A}_\kappa\left(\hat{\omega}\right)\right) + \frac{q}{q-1}\frac{p(q)-1}{p(q)}\left(S_{p(q)}\left(\hat{\rho}\right)-S_{p(q)}\left(\hat{\omega}\right)\right)\;,
\end{align}
where we have used in sequence \eqref{eq:propSa}, \eqref{eq:defSa}, \eqref{eq:pq} and \eqref{eq:defSa} again.
Since $S(\hat{\rho})=S(\hat{\omega})$, we have from \eqref{eq:propSa}
\begin{equation}
\lim_{p\to1}(S_p(\hat{\rho})-S_p(\hat{\omega}))=0\;,\quad\lim_{q\to 1}S_q(\mathcal{A}_\kappa(\hat{\omega}))=S(\mathcal{A}_\kappa(\hat{\omega}))\;.
\end{equation}
Then, the claim \eqref{CLAIM} follows taking the limit $q\to1$ in~\eqref{eq:SAq} and using that
\begin{equation}
\lim_{q\to 1}p(q)=1\quad \text{and} \quad 0\le \frac{q}{q-1}\frac{p(q)-1}{p(q)}\le1\;.
\end{equation}
\end{proof}
\end{thm}
The proof of the CMOE conjecture for an arbitrary one-mode phase-covariant quantum Gaussian channel can be obtained by merging
Theorem~\ref{thm:SP} and Theorem~\ref{thm:SA}.
\begin{thm}[CMOE conjecture for phase-covariant quantum Gaussian channels]\label{thm:main}
Gaussian thermal input states minimize the output entropy of any one-mode phase-covariant quantum Gaussian channel among all the input states with a given entropy, i.e.\ for any $0\le\lambda\le1$, $\kappa\ge1$, $E\ge0$ and any quantum state $\hat{\rho}$
\begin{align}
S\left(\mathcal{E}_{\lambda,E}\left(\hat{\rho}\right)\right) &\ge g\left(\lambda\;g^{-1}\left(S\left(\hat{\rho}\right)\right)+\left(1-\lambda\right)E\right)\;,\\
S\left(\mathcal{N}_E\left(\hat{\rho}\right)\right) &\ge g\left(g^{-1}\left(S\left(\hat{\rho}\right)\right)+E\right)\;,\\
S\left(\mathcal{A}_{\kappa,E}\left(\hat{\rho}\right)\right) &\ge g\left(\kappa\;g^{-1}\left(S\left(\hat{\rho}\right)\right)+\left(\kappa-1\right)\left(E+1\right)\right)\;.
\end{align}
\begin{proof}
We have from \eqref{eq:dec} and Theorem \ref{thm:SA}
\begin{align}
S\left(\mathcal{E}_{\lambda,E}\left(\hat{\rho}\right)\right) &= S\left(\mathcal{A}_{\kappa'}\left(\mathcal{E}_{\lambda'}\left(\hat{\rho}\right)\right)\right)\nonumber\\
&\ge g\left(\kappa'\;g^{-1}\left(S\left(\mathcal{E}_{\lambda'}\left(\hat{\rho}\right)\right)\right)+\kappa'-1\right)\;.
\end{align}
Since $g$ is increasing, also $g^{-1}$ is increasing, and $S\mapsto g(\kappa'g^{-1}(S)+\kappa'-1)$ is increasing, too.
Then, Theorem \ref{thm:SP} implies
\begin{equation}
S\left(\mathcal{E}_{\lambda,E}\left(\hat{\rho}\right)\right) \ge g\left(\kappa'\lambda'\;g^{-1}\left(S\left(\hat{\rho}\right)\right)+\kappa'-1\right)\;,
\end{equation}
i.e. the claim.

The proofs for $\mathcal{N}_E$ and $\mathcal{A}_{\kappa,E}$ are identical.
\end{proof}
\end{thm}
The proof of the CMOE conjecture for phase-contravariant channels follows from an observation in \cite{qi2016thermal}, Section VI.
\begin{thm}[CMOE conjecture for phase-contravariant quantum Gaussian channels]
Gaussian thermal input states minimize the output entropy of any one-mode phase-contravariant quantum Gaussian channel among all the input states with a given entropy, i.e.\ for any $\kappa\ge1$, $E\ge0$ and any quantum state $\hat{\rho}$
\begin{equation}
S\left(\tilde{\mathcal{A}}_{\kappa,E}\left(\hat{\rho}\right)\right) \ge g\left(\left(\kappa-1\right)\left(g^{-1}\left(S\left(\hat{\rho}\right)\right)+1\right)+\kappa\;E\right)\;.
\end{equation}
\begin{proof}
(\cite{qi2016thermal}, Section VI) The claim follows from Theorem \ref{thm:main} observing that any phase-contravariant channel can be decomposed as a phase-covariant channel followed by the transposition, that does not change the entropy.
\end{proof}
\end{thm}

\section{Conclusions}
We have proved that Gaussian thermal input states minimize the output von Neumann entropy of one-mode phase covariant and contravariant quantum Gaussian channels among all the input states with a given entropy.
This result finally permits to extend the Entropy Power Inequality to the quantum regime, and to prove the optimality of Gaussian encodings for both the triple trade-off coding and broadcast communication with the quantum-limited amplifier \cite{qi2016capacities}.
The future challenge is the extension of our result to the multimode scenario.
Our proof has exploited a new technique that can be used to determine the minimum output entropy for fixed input entropy for any quantum channel whose $p\to q$ norms are known.

\section*{Acknowledgements}
GdP acknowledges financial support from the European Research Council (ERC Grant Agreement no 337603), the Danish Council for Independent Research (Sapere Aude Grant no 1323-00025B) and VILLUM FONDEN via the QMATH Centre of Excellence (Grant No. 10059).

\begin{widetext}
\appendix
\section{}\label{app}
\begin{lem}\label{lem}
For any $\kappa>1$, any $S>0$ and any $1<q<3/2$ there exists $1<p<q$ such that the $p\to q$ norm of $\mathcal{A}_\kappa$ is saturated by the thermal Gaussian state with entropy $S$.
\begin{proof}
From Theorem \ref{thm:pq} the $p\to q$ norm of $\mathcal{A}_\kappa$ is saturated by some thermal Gaussian state.
It is then sufficient to prove that we can choose $p$ such that this state has entropy $S$.
We use a different parametrization of thermal Gaussian states, with
\begin{align}
z &:=\frac{E}{E+1}\;,\qquad 0\le z<1\;,\\
\hat{\omega}_z &=\sum_{n=0}^\infty\left(1-z\right)z^n\;|n\rangle\langle n|\;.
\end{align}
The transformation rule for $z$ following from Eq. \eqref{eq:AE'} is $\mathcal{A}_\kappa\left(\hat{\omega}_z\right)=\hat{\omega}_{z'}$, with
\begin{equation}\label{eq:z'}
z'=\frac{z+\kappa-1}{\kappa}\;.
\end{equation}

Let $0<\bar{z}<1$ be such that $\hat{\omega}_{\bar{z}}$ has entropy $S$.
The claim follows if we prove that we can choose $p$ such that the function
\begin{equation}\label{eq:rate}
z\mapsto\frac{\left\|\mathcal{A}_\kappa\left(\hat{\omega}_z\right)\right\|_q}{\left\|\hat{\omega}_z\right\|_p}\;,\qquad 0<z<1
\end{equation}
has a global maximum at $z=\bar{z}$.
We can compute
\begin{align}
\ln\left\|\hat{\omega}_z\right\|_p &= \ln\left(1-z\right)-\frac{1}{p}\ln\left(1-z^p\right)\;,\nonumber\\
\frac{\partial}{\partial z}\ln\left\|\hat{\omega}_z\right\|_p &= -\frac{1-z^{p-1}}{\left(1-z\right)\left(1-z^p\right)}\;.
\end{align}
For any $0<z<1$, let $z'$ be such that $\mathcal{A}_\kappa\left(\hat{\omega}_z\right)=\hat{\omega}_{z'}$.
Similarly, let $\bar{z}'$ be such that $\mathcal{A}_\kappa\left(\hat{\omega}_{\bar{z}}\right)=\hat{\omega}_{\bar{z}'}$.
With the help of \eqref{eq:z'} we have
\begin{equation}\label{eq:derphi}
\frac{\partial}{\partial z}\ln\frac{\left\|\mathcal{A}_\kappa\left(\hat{\omega}_z\right)\right\|_q}{\left\|\hat{\omega}_z\right\|_p} = \frac{\phi(z,p)-\phi(z',q)}{1-z}\;,
\end{equation}
where
\begin{equation}
\phi(z,p):=\frac{1-z^{p-1}}{1-z^p}\;.
\end{equation}

Let us first prove that we can choose $1<p<q$ such that
\begin{equation}\label{eq:der}
\left.\frac{\partial}{\partial z}\ln\frac{\left\|\mathcal{A}_\kappa\left(\hat{\omega}_z\right)\right\|_q}{\left\|\hat{\omega}_z\right\|_p}\right|_{z=\bar{z}}=0\;,
\end{equation}
i.e. $\phi(\bar{z},p)=\phi(\bar{z}',q)$.
The function $p\mapsto\phi(\bar{z},p)$ is continuous for $p\ge1$.
The claim then follows if we prove that
\begin{equation}\label{eq:phipq}
\phi(\bar{z},1)<\phi(\bar{z}',q)<\phi(\bar{z},q)\;.
\end{equation}
The first inequality in \eqref{eq:phipq} follows since $\phi(\bar{z},1)=0<\phi(\bar{z}',q)$.
The second inequality in \eqref{eq:phipq} follows since $\bar{z}'>\bar{z}$ and $\phi(z,q)$ is decreasing in $z$.
This last property holds since
\begin{equation}
\frac{\partial}{\partial z}\ln\phi(z,p) = \frac{z^{p-2}}{1-z^{p-1}}\frac{1+p\left(z-1\right)-z^p}{1-z^p}<0\;.
\end{equation}

There remains to prove that any $\bar{z}$ satisfying \eqref{eq:der} is a global maximizer for \eqref{eq:rate}.
For $z=0$ we have $z'=1-1/\kappa>0$, hence
\begin{equation}
\phi(z',q)<\phi(0,q)=1=\phi(0,p)\;,
\end{equation}
and
\begin{equation}
\left.\frac{\partial}{\partial z}\ln\frac{\left\|\mathcal{A}_\kappa\left(\hat{\omega}_z\right)\right\|_q}{\left\|\hat{\omega}_z\right\|_p}\right|_{z=0}>0\;.
\end{equation}
Then, $z=0$ cannot be a maximizer.
Moreover,
\begin{equation}
\lim_{z\to1}\phi(z,p)=1-\frac{1}{p}<1-\frac{1}{q}=\lim_{z\to1}\phi(z,q)\;,
\end{equation}
hence
\begin{equation}
\lim_{z\to 1}\frac{\partial}{\partial z}\ln\frac{\left\|\mathcal{A}_\kappa\left(\hat{\omega}_z\right)\right\|_q}{\left\|\hat{\omega}_z\right\|_p}=-\infty\;,
\end{equation}
and the supremum cannot be achieved for $z\to1$.
The maximizer must then be in the open interval $(0,1)$, and the claim follows if we prove that \eqref{eq:derphi} vanishes only for $z=\bar{z}$.
Equivalently, we will prove that $z\mapsto\phi(z,p)/\phi(z',q)$ is strictly decreasing.
We have
\begin{equation}
\frac{\partial}{\partial z}\ln\frac{\phi(z,p)}{\phi(z',q)} = \frac{f(z',q)-f(z,p)}{1-z}\;,
\end{equation}
where
\begin{equation}
f(z,p):=z^{p-2}\frac{1-z}{1-z^{p-1}}\left(p\frac{1-z}{1-z^p}-1\right)\;.
\end{equation}
We must then prove that
\begin{equation}\label{eq:fpq}
f(z',q)<f(z,p)\;.
\end{equation}
We recall that $1<p<q<3/2$, hence $z\mapsto z^{p-1}$ is concave, and
\begin{equation}
\frac{\partial}{\partial z}\left(z^{p-2}\frac{1-z}{1-z^{p-1}}\right) = z^{p-1}\frac{\left(p-1\right)\left(1-z\right)+z^{p-1}-1}{\left(z-z^p\right)^2}\le0\;.
\end{equation}
It follows that $z\mapsto z^{p-2}\frac{1-z}{1-z^{p-1}}$ is positive and decreasing.
Since $z\mapsto z^p$ is convex,
\begin{equation}
z\mapsto \psi(z,p):=p\frac{1-z}{1-z^p}-1
\end{equation}
is decreasing.
Since $\lim_{z\to1}\psi(z,p)=0$, $\psi$ is also positive.
Then, $z\mapsto f(z,p)$ is decreasing.
Since $z'>z$, we have
\begin{equation}
f(z',q)-f(z,p)\le f(z,q)-f(z,p)\;.
\end{equation}
Since $p<q$, the claim \eqref{eq:fpq} follows if we prove that $p\mapsto f(z,p)$ is decreasing.
We have
\begin{align}
&\frac{\partial}{\partial p}f(z,p)=\frac{z^{p-2}\left(1-z\right)^2}{\left(1-z^p\right)^2\left(1-z^{p-1}\right)^2}\nonumber\\
&\times\left(\left(1-z^{p-1}\right)\left(1-z^p\right)+\left(1-z^{2p-1}\right)\ln z^{p-1}-\frac{z\ln z}{1-z}\left(1-z^{p-1}\right)^2\right)\;.
\end{align}
We need the following inequalities:
\begin{align}\label{eq:in1}
-\frac{z\ln z}{1-z} &< \sqrt{z}\;,\\
\ln z^{p-1} &< z^{p-1}-1-\frac{\left(1-z^{p-1}\right)^2}{2}\;.\label{eq:in2}
\end{align}
\eqref{eq:in1} follows from $t<\sinh t$ with $t=-\frac{\ln z}{2}>0$.
\eqref{eq:in2} follows from the Taylor series
\begin{equation}
\ln\left(1-x\right)=-\sum_{n=1}^\infty\frac{x^n}{n}
\end{equation}
with $x=1-z^{p-1}>0$.
We then get
\begin{equation}
\frac{\partial}{\partial p}f(z,p)<-\frac{z^{p-2}\left(1-z\right)^2\left(1-z^{p-\frac{1}{2}}\right)\left(1-2z^\frac{1}{2}+z^{p-\frac{1}{2}}\right)}{2\left(1-z^p\right)^2}\;.
\end{equation}
Since $p<3/2$, we have
\begin{equation}
1-2z^\frac{1}{2}+z^{p-\frac{1}{2}}>\left(1-\sqrt{z}\right)^2>0\;,
\end{equation}
then $\frac{\partial}{\partial p}f(z,p)<0$ and the claim follows.
\end{proof}
\end{lem}
\end{widetext}

\bibliography{biblio}
\bibliographystyle{apsrev4-1}
\end{document}